# Distinguishing articles in questionable and non-questionable journals using quantitative indicators associated with quality

Dimity Stephen
stephen@dzhw.eu, ORCID: 0000-0002-7787-6081
German Centre for Higher Education Research and Science Studies
Schützenstraße 6a, Berlin, Germany, 10117

**Abstract** This study investigates the viability of distinguishing articles in questionable journals (QJs) from those in non-QJs on the basis of quantitative indicators typically associated with quality. Subsequently, I examine what can be deduced about the quality of articles in QJs based on the differences observed. I contrast the length of abstracts and full-texts, prevalence of spelling errors, text readability, number of references and citations, the size and internationality of the author team, the documentation of ethics and informed consent statements, and the presence erroneous decisions based on statistical errors in 1,714 articles from 31 QJs, 1,691 articles from 16 journals indexed in Web of Science (WoS), and 1,900 articles from 45 mid-tier journals, all in the field of psychology. The results suggest that QJ articles do diverge from the disciplinary standards set by peer-reviewed journals in psychology on quantitative indicators of quality that tend to reflect the effect of peer review and editorial processes. However, mid-tier and WoS journals are also affected by potential quality concerns, such as under-reporting of ethics and informed consent processes and the presence of errors in interpreting statistics. Further research is required to develop a comprehensive understanding of the quality of articles in QJs.

**Keywords**: article quality; questionable journals; scientometric indicators; Web of Science; Cabell's

## 1. Introduction

Concern has been growing in the academic community over the last several years about the emergence of questionable journals (QJs). QJs – also known as predatory journals – exhibit questionable publishing practices, such as bypassing peer review, soliciting researchers for submissions, or falsely claiming indexation or impact metrics (e.g., Oermann et al., 2018). While there is still much debate in the field about how to classify journals as questionable (Krawczyk & Kulczycki, 2021a), the concerns about their impact on academic discourse are more well-established. Perhaps this is justified: if QJs do simply publish all research that crosses their editorial desks, then they could have the potential to contaminate academic discourse with harmful information (Oermann et al., 2018), waste animals' lives and human participants' time (Moher et al., 2017), misdirect $75 million of funding annually to publishing fees for low-visibility outputs (Linacre, 2022), tarnish the reputation of institutions, researchers (Kulczycki & Rotnicka, 2022) and all open access journals (Linacre, 2022), and contribute to diminishing public trust in science.

However, we currently lack the necessary evidence regarding the quality of research in questionable journals to wholly justify our concerns. Our current characterisation of questionable journals largely ignores the complex geopolitical relations at play and disproportionately disadvantages countries in the semi-periphery of the global academic system (Krawczyk & Kulczycki, 2021a). We must also consider researchers' motivations for utilising QJs, which tend to stem from publication pressure and inexperience, rather than finding a venue for intentionally subpar work (Frandsen, 2019). Further, empirical studies of QJs' practices, content and quality constitute only 19% of the existing literature on this topic (Krawczyk & Kulczycki, 2021b), suggesting we lack empirical evidence about the quality of research in QJs. We are thus arguably operating in a state of moral panic: we have identified a particular group as a threat to the standards of our community, and, while this threat may be real, we do not currently have sufficient justification for our intense concerns (Cohen, 1972). Consequently, further research is required to investigate the quality of articles in QJs to understand their potential impact on the academic and broader communities. However, manual review of QJs is resource-intensive. As such, I assess in this study whether articles in QJs are distinguishable from those in non-QJs based on quantitative indicators typically associated with quality, and investigate what can be deduced about the quality of research in QJs from any observable differences.

*1.1. Defining and measuring article quality*

What constitutes research quality is a frequently debated topic. However, Langfeldt et al (2020) identified three dimensions that continuously emerged in studies of research quality: originality or novelty, plausibility or reliability, and value or utility. Originality refers to the ability of a work to progress a field via new information, whether theoretical, methodological, small- or large-scale. Plausibility reflects the rigour of a work in its methods, theoretical grounding, ethics, and clarity. Value refers to a work's academic value in contributing information to its field and also its broader societal value (Langfeldt et al., 2020). The framework also distinguishes between field-type and space-type notions of quality. Field-type notions emerge directly from the research community and are based on their communally accepted knowledge sources and practices. Judgements about these notions of quality occur via peer review and focus on the standard of the content and whether it serves to extend the community's current knowledge. In contrast, space-type quality notions are derived from knowledgeable laypersons outside of the community, such as policy-makers and funding bodies. Here, quality is often considered in relation to the societal and economic outputs of research and may be assessed without peer review via proxies that represent the underlying construct, e.g., journal reputation and citation indicators. Research quality may thus be defined as these three core facets and conceptualised differently depending on the community's notion of quality.

The coexistence of these two contrasting notions of research quality has led to the use of a wide array of indicators to assess this construct in contemporary research, as shown in Table 1. Here, I have divided the indicators into five overarching themes: theory, methodology, communication, impact, and collaboration. Note that this table pertains to article-level indicators and excludes author- and journal-level indicators, such as the h-index or Journal Impact Factor. Under Langfeldt et al's (2020) framework, the majority of these indicators pertain to plausibility: the rigour of the methodology, theoretical grounding, and communication. Only the "clear contribution to field made" indicator assesses the work's originality, and also its value or utility. The latter construct is also measured via the proxies of citation and engagement metrics.

Indeed, citation metrics are the most commonly applied proxy of article quality (Kousha & Thelwall, 2024). This metric is usually defined as the number of citations an article has received within a specific time-period, often normalised for the article's field to facilitate comparisons. More highly cited articles are inferred to be of higher quality due to their utility to their fields, as evidenced by other authors citing them (Kousha & Thelwall, 2024). Citation-based indicators are widely applied in, for instance, national research evaluation exercises and in ranking articles, journals, researchers, and institutions. However, advocates for responsible evaluation practices urge moving on from using citations in research assessment, arguing they provide an incomplete picture of research quality (Coalition for Advancing Research Assessment, 2023). Indeed, citations correlated only weakly with peer-assessed article quality in the humanities and social sciences ($r$=0.1-0.3) and moderately in the medical and life sciences and engineering fields ($r$=0.3-0.5), using data from the United Kingdom's Research Evaluation Framework (REF; Thelwall et al., 2023a). This suggests citation impact captures just one aspect of article quality as a proxy for its academic value.

More recently, engagement metrics have been used as an alternative or supplement to citations by examining the uptake of the paper both inside and outside of academia as a proxy for academic and societal value. These "altmetrics" include interactions with the article on social media, bookmarking in reference managers, downloads of the article from journal websites, and mainstream media coverage. Altmetrics tend to reflect the interest of the general public in an article and thus vary by discipline, with greater interest in the social sciences and humanities than natural sciences and engineering (Haustein, Costas, & Larivière, 2015). Only weak to moderate associations have been found between Mendeley readers, and Facebook, Twitter, news and blog activity with peer review scores from the UK's 2021 REF (Thelwall et al., 2023b). Altmetrics may thus embody a space-type quality notion and align only loosely with field-type notions of research quality.

Several articles have also distilled quality into a set of indicators based on the presence or absence of particular theoretical, methodological, and communication aspects that largely reflect plausibility or rigour. For instance, regarding theoretical aspects of the work, these indicators pertain to a sufficient review of prior literature, establishing a clear theoretical framework for the study, adequately acknowledging the study's limitations, and drawing valid conclusions based on the study's findings (Bordage, 2001; Goodman, 1994; Henly & Dougherty, 2009; Heuritsch, 2021; McCutcheon et al., 2016; Nieminen, Carpenter, Rucker & Schumacher, 2006; Oermann et al., 2018). Methodologically, these indicators include the absence of statistical errors, the selection of a sufficient sample size, and the application of appropriate methods to answer the research question (Bordage, 2001; Goodman, 1994; Heuritsch, 2021; McCutcheon et al., 2016; Nieminen et al., 2006; Oermann et al., 2018). Compliance with reporting guidelines and the documentation of ethics review processes were also regarded as signs of high quality in medicine (Carneiro et al., 2020; Bianchini et al. (2020); Goodman, 1994; Oermann et al., 2018). Similarly, the clarity of the text in describing its purpose, methods, results, and its connection to its field demonstrates its plausibility in relation to communication.

Indicators about communication also encompass two other factors. First, the structure of the text and its alignment with the field's expected practices, such as an abstract's presence and length, the length of the article full-text, the number of (effective) figures/tables used, and whether the article follows a logical structure (Bordage, 2001; Fronzetti Colladon, D'Angelo, & Gloor, 2020; Goodman, 1994; Henly & Dougherty, 2009; Kousha & Thelwall, 2024; Oermann et al., 2018; Song, Chen, & Zhao, 2023). The second factor pertains to the characteristics of the text itself, such as spelling or grammar mistakes or plagiarism, and the text's sentiment, lexical diversity, lexical density, complexity, commonness, and readability (Ante, 2022; Bordage, 2001; Chen et al., 2020; Fronzetti Colladon et al., 2020; Jin et al., 2021; Lei & Yan, 2016; Lu et al., 2019; McCutcheon et al., 2016; Owen & Nichols, 2019; Song et al., 2023). Sentiment refers to the positive or negative tone of the article (Fronzetti Colladon et al., 2020), usually measured by the use of pre-defined positive or negative words. Lexical diversity is the ratio of unique word stems to the total number of words in a text, with higher scores indicating a more diverse vocabulary (Fronzetti Colladon et al., 2020). Lexical density is the ratio of word types, such as nouns and verbs, to the total words in a text, with more dense texts expected to carry more information (Song et al., 2023; Chen et al., 2020). Complexity measures the variety in the frequency of the words used in a text, with higher scores inferring greater complexity and perhaps the presentation of more complex ideas (Fronzetti Colladon et al., 2020). Commonness compares the words used in a text against a similar corpus to measure the distinctiveness of the text (Fronzetti Colladon et al., 2020), thus a potential indicator also of originality. Readability can be measured using various methods, but fundamentally reflects how easily the text can be read and understood (Jin et al., 2021).

Unlike some theoretical and methodological aspects of an article, where it is relatively clear that quality has been achieved when the stated criteria have been fulfilled, in the case of some communication characteristics, there is greater ambiguity about what constitutes quality. For instance, using citations as a proxy of quality, negligible but significant associations were observed between citations and sentiment ($r$=0.05), lexical diversity (0.03), and commonness (-0.02) in a sample of 224,000 chemical engineering abstracts, but no association with lexical complexity was found (0.006; Fronzetti Colladon et al., 2020). In library and information science (LIS), articles that were more lexically dense received more citations and were more often published in top quartile journals (Song et al., 2023). Conversely, chemical engineering articles with more positive sentiment appeared more often in lower ranked journals (Fronzetti Colladon et al., 2020). Lexical density and diversity were also largely unrelated to the number of times 63,000 articles in PLoS journals were viewed or downloaded (Chen et al., 2020), or articles' Altmetrics Attention Scores (Jin et al., 2021). The influence of readability is particularly complicated as lower readability may stem either from poor writing or from the conveyance of complex ideas. For instance, difficult texts were more likely to be rejected from medical conferences (Bordage, 2001) and more readable texts received more citations in one LIS study (Song et al., 2023), although another LIS study found readability and citations were unrelated (Lei & Yan, 2016). Conversely, the most highly cited articles in computer science topics were also the most unreadable (Ante, 2022). The use of lexical characteristics to assess article quality is an emerging field with only a handful of studies to date that have found mixed associations with citations, altmetrics, and journal rankings as quality proxies.

One final aspect – collaboration – is sometimes used to measure quality. Here, the number of co-authors on a paper (e.g., Fronzetti Colladon et al., 2020) or whether the paper is an international collaboration (Salimi, 2017) is assessed, with collaboration typically interpreted as higher quality. However, there are important disciplinary differences to consider, such as the standard practice in physics to publish articles with hundreds or even thousands of co-authors, and the humanities often focus on nationally-oriented topics, thus reducing international partnerships. To this point, author counts were recently observed to moderately correlate with peer-assessed article quality in the life and physical sciences ($r$=0.2-0.4). However, no associations were found in engineering, social sciences and humanities (Thelwall et al., 2023c), although a moderate association of 0.22 in chemical engineering has been observed elsewhere (Fronzetti Colladon et al., 2020). A recent meta-analysis also found only a weak, positive association ($r$=0.15) between collaboration and citations (Shen et al., 2021). In cases such as collaboration and communication, where the association with common proxies of quality is unclear, it may be more informative to directly compare these characteristics of different groups of articles against one another to identify deviations in practice, rather than via quality proxies.

Table 1. Indicators of article quality used in previous studies

| Quality indicator | Relevant studies |
|---|---|
| **Theory** | |
| 1. Sufficient review of the literature* | Henly & Dougherty (2009); Oermann et al. (2018); McCutcheon et al. (2016) |
| 2. Clear theoretical framework established | Henly & Dougherty (2009) |
| 3. Clear contribution to field made | Oermann et al. (2018); Heuritsch (2021); McCutcheon et al. (2016); Bordage (2001) |
| 4. The research purpose or question is identified | Goodman (1994); Henly & Dougherty (2009); Oermann et al. (2018); Nieminen et al. (2006); Bordage (2001) |
| 5. Appropriate discussion of limitations | Goodman (1994); Bordage (2001) |
| 6. Valid conclusions are drawn based on results | Henly & Dougherty (2009); Oermann et al. (2018) |
| **Methodology** | |
| 7. Absence of statistical errors* | McCutcheon et al. (2016) |
| 8. Appropriateness of the subjects and sample size | Oermann et al. (2018); Bordage (2001) |
| 9. Appropriateness of the methods | Oermann et al. (2018); Heuritsch (2021); Nieminen et al. (2006); Bordage (2001) |
| 10. Adequate description of methods | Goodman (1994); Heuritsch (2021); Nieminen et al. (2006) |
| 11. Adequate reporting of results | Goodman (1994); Heuritsch (2021); Nieminen et al. (2006); Bordage (2001) |
| 12. Compliance with reporting guidelines | Carneiro et al. (2020); Goodman (1994); Bianchini et al. (2020) |
| 13. Documentation of ethics review* | Oermann et al. (2018) |
| **Communication** | |
| 14. Presence of an abstract and its length* | Fronzetti Colladon et al. (2020); Oermann et al. (2018); Song et al., (2023) |
| 15. Logical organisation of the article | Goodman (1994); Henly & Dougherty (2009); Oermann et al. (2018); Bordage (2001) |
| 16. Article length* | Oermann et al. (2018); Song et al., (2023) |
| 17. Number of (effective) figures/tables used | Bordage (2001); Oermann et al. (2018) |
| 18. Absence of spelling or grammar mistakes* | McCutcheon et al. (2016) |
| 19. Absence of plagiarism | Owens & Nicoll (2019) |
| 20. Sentiment | Fronzetti Colladon et al. (2020) |
| 21. Lexical diversity | Fronzetti Colladon et al. (2020); Song et al. (2023); Chen et al. (2020) |
| 22. Lexical density | Song et al. (2023); Chen et al. (2020); Jin et al. (2021) |
| 23. Complexity | Fronzetti Colladon et al. (2020); Lu et al. (2019); Jin et al. (2021) |
| 24. Commonness | Fronzetti Colladon et al. (2020) |
| 25. Readability* | Ante (2022); Bordage (2001); Lei & Yan (2016); Song et al. (2023) |
| **Impact** | |
| 26. Citation metrics* | e.g., Thelwall et al., (2023a) |
| 27. Engagement metrics | e.g., Thelwall et al., (2023b) |
| **Collaboration** | |
| 28. Collaboration* | Fronzetti Colladon et al. (2020), Salimi (2017), Thelwall et al. (2023c) |

\* denotes the indicators that are used in the current study

*1.2. Previous studies of the quality of articles in QJs*

This overview highlights the diversity of indicators of quality and the profusion of research in this area. However, very little of this research has been applied to study the quality of articles in QJs; I identified only five studies that had compared QJs to their non-questionable counterparts in an attempt to identify variations in quality. First, a comparison of 410 randomised control trials (RCTs) in physical therapy found that articles in QJs had significantly poorer compliance with reporting guidelines than articles in non-QJs (Bianchini et al., 2020). Yan et al (2018) similarly observed lower reporting quality in 3 orthopaedics RCTs in QJs compared to 6 RCTs in non-QJs, although the small sample size limits generalisability. In a blind comparison of 25 articles each from non/QJs in psychology, raters scored articles in QJs significantly more poorly on measures of the theoretical grounding, methodology, and presentation (McCutcheon et al., 2016). Using similar criteria, 48.4% of 353 articles in questionable nursing journals were rated as poor quality and 5% of articles contained information that could be potentially harmful to patients. However, 47.9% of articles were of average quality and 3.7% were rated as excellent (Oermann et al., 2018). Finally, in a previous study I found that 348 articles in questionable medical journals were less often international collaborations and less often cited than similar articles in non-QJs. However, QJ articles were still widely cited in mainstream journals, perhaps suggesting they were of acceptable quality (Stephen, 2023). As such, while this limited number of studies shows articles in QJs tend to perform more poorly than non-QJ articles, there is still much to be clarified regarding the quality of QJs.

*1.3. Aims of the current study*

In this study, I compare indicators of quality between samples sourced from i) QJs, ii) non-QJs indexed in the Web of Science (WoS), and iii) non-QJs not indexed in WoS. These categories represent a potential spectrum of quality: journals in WoS have been directly selected due to their high quality and centrality to their fields; non-WoS and non-QJs represent a middle tier of journals that are neither questionable nor top-tier; and QJs as potentially lower quality due to the classification of their publishing venue as questionable. The two non-QJ samples thus constitute a baseline for the characteristics of articles published in peer-reviewed journals that we can use to assess whether articles in QJs deviate from this standard. Importantly, I examine articles in only one field of research, to ensure field-specific practices in collaboration, methodology and communication are controlled. However, the manual review of studies required to qualitatively assess studies is resource-intensive, which likely contributes to the dearth of studies on this topic. Consequently, here I examine quality via quantifiable indicators. The study's research questions are thus: i) are articles in QJs distinguishable from those in non-QJs based on quantitative indicators of quality, and ii) what can consequently be deduced about the quality of articles in QJs from any differences observed?

The indicators selected to be examined are: i) a sufficient review of the literature, operationalised as the number of references, ii) the length of the abstract, iii) the length of the article, iv) the presence of spelling mistakes, v) the readability of the article, vi) the number of citations received, vii) the size and internationality of the author team, viii) documentation of ethics reviews and obtaining informed consent, and ix) the presence of statistical errors, operationalised as misreporting of p-values. This selection sampled indicators from each category of theory, methodology, communication, impact, and collaboration that were able to be operationalised using quantitatively measureable variables.

**3. Methods**

An overview of the method used to collect and analyse the study's data, the sources used, and the inclusion criteria applied are shown in Figure 1.

*3.1. Sample selection and collection*

I defined QJs as those listed in Cabell's Predatory Reports (CPR). CPR is a subscription-based service that classifies journals as "predatory" via manual review of the journal against more than 60 criteria, such as publishing practices and accurate use of metrics[1]. I selected CPR as the basis for identifying QJs as it is current and transparent in its justification for classifying journals as questionable. To identify in-scope journals, I first extracted the title and ISSNs of all journals that published articles between 2010 and 2020 from Dimensions as it contains a broader sample of research than the curated databases (Visser et al., 2021). Dimensions data were obtained from Digital Science by the Kompetenznetzwerk Bibliometrie (KB)[2] as a snapshot of data up to April 2021. As the process to identify journals in CPR was largely manual, I first reduced the sample of journals by excluding those in the Directory of Open Access Journals (DOAJ) or WoS (KB's in-house April 2022 snapshot) as these journals were unlikely to be questionable due to the screening processes for inclusion in these indexes. I then searched for the remaining journals' titles/ISSNs in CPR to identify those classified as questionable as of 16 August 2022.

---

[1] https://cabells.com/
[2] https://bibliometrie.info/en/

It was important that the journals in each sample were from the same discipline to account for any differences in article characteristics between disciplines. Consequently, I restricted the samples to journals with "psycholo*" in the title, capturing those in the field of psychology. Further, I required that the article's full-text was available for examination. Unpaywall[3] is a publicly available database of Open Access (OA) content from more than 50,000 publishers that facilitates access to scholarly content. I matched in-scope articles from Dimensions and WoS with KB's snapshot of Unpaywall as of July 2022 via DOI and then extracted the URL to articles' full-texts. As such, the inclusion criteria applied to all samples were i) an article published between 2010 and 2020 ii) in a journal with "psycholo*" in the title and iii) a URL to a full-text in Unpaywall. Further, the QJ sample's journals were classified as "predatory" by CPR, the WoS sample's journals were indexed in WoS and OA, and the mid-tier (non-WoS/non-QJ) sample's journals were indexed in the DOAJ. The OA inclusion criterion was included for the two non-QJ samples to account for any citation (dis)advantage as all QJs were OA.

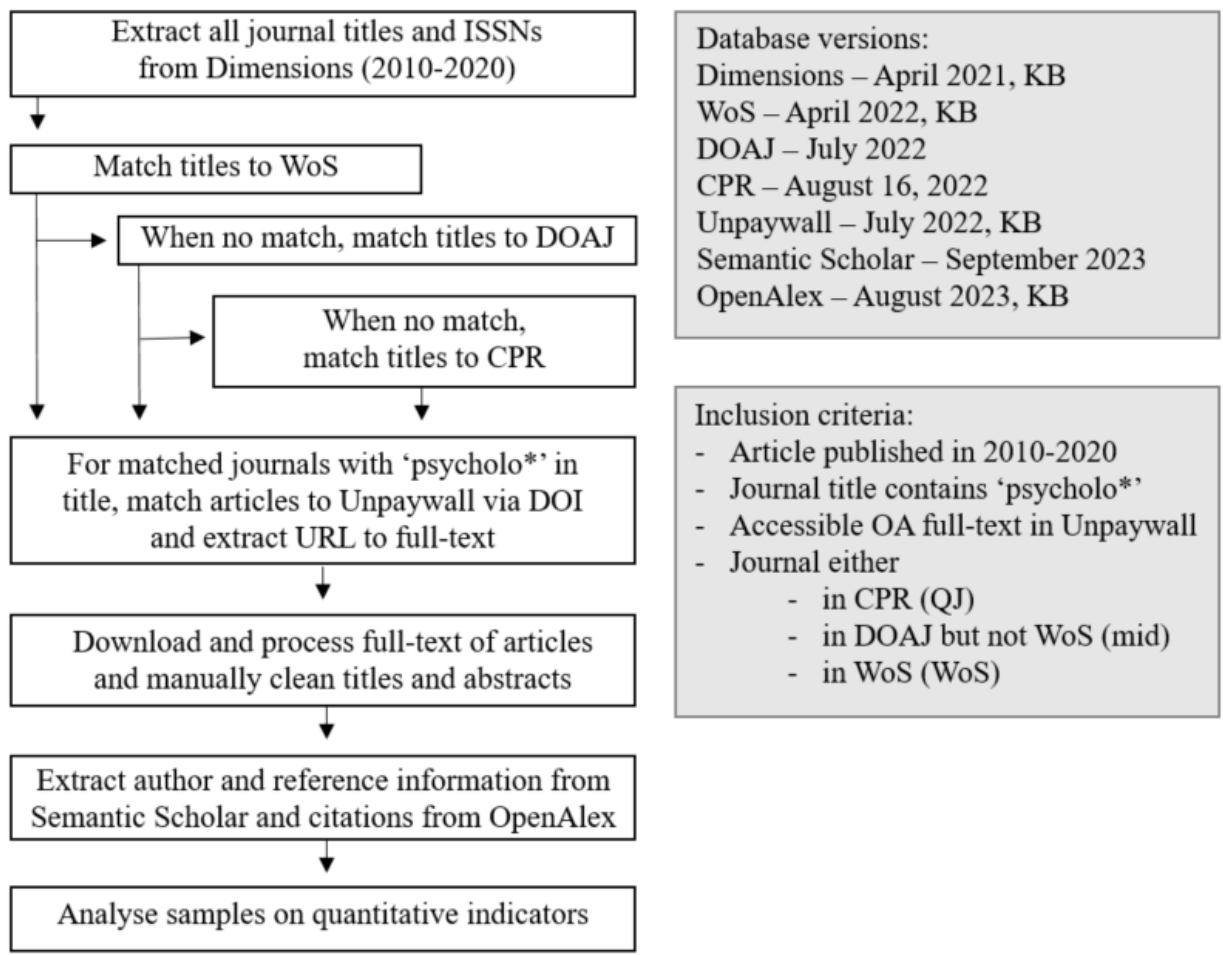

**Fig. 1** The method used to collect and analyse data, its sources, and inclusion criteria

I then extracted the journal title, article title, and the URL to the full-text from Unpaywall for all articles within scope. As I could not filter to only the "article" document type in either Unpaywall or Dimensions, I excluded as much non-research material as possible by removing documents with keywords in the titles such as "editorial", "book review", "letter to/from the editor", "correction", "opinion", etc. I also removed articles published in languages other than English based on titles to ensure consistency in calculating language-based indicators. I used the tidyverse package (Wickham et al., 2019) and base functions in R (R Core Team, 2023) for this processing.

---

[3] https://unpaywall.org/

*3.2. Data extraction and cleaning*

I downloaded a PDF version of the full-text of each article using the download.file function of the utils package in R (R Core Team, 2023). Articles that timed out or were unavailable at this URL were excluded. I then converted the PDFs to XML format via the open source software Content ExtRactor and MINEr (CERMINE; Tkaczyk et al., 2015), as this format is able to be read by the JATSdecoder R package (Böschen, 2022). Using JATSdecoder I extracted the articles' titles, abstracts, references, and full-texts. I manually examined the extracted titles and abstracts and corrected any missing entries or misspellings introduced by issues reading PDFs. During this cleaning, I also excluded any additional articles identified as non-research material or written in languages other than English. For articles with abstracts in more than one language, I retained the English version. I did not undertake any additional cleaning for the full-texts due to resource constraints.

*3.3. Analysis of quantitative indicators*

The methods used to analyse each of the specific indicators are described in the following subsections. I undertook all analyses in R, with the relevant packages noted in each subsection. To examine the groups for statistical differences, I used one-way ANOVAs to compare all groups, unless otherwise stated. Where significant differences were observed, I used Tukey's HSD pairwise comparisons to identify which groups differed. The graphs were produced using ggplot2 (Wickham et al., 2019) and patchwork (Pedersen, 2022).

*3.3.1. Length of abstract and full-text*

I calculated the length of the abstracts and full-texts of each article as the number of words contained in each using stringr in the tidyverse package (Wickham et al., 2019). The full-text of the article does not include the reference list, but does include end matter, such as acknowledgements, and ethics and funding statements.

*3.3.2. Percentage of articles with misspellings*

I identified misspelt words in the titles and abstracts using the hunspell package (Ooms, 2023). I considered misspellings as those words where the intended word was discernible but misspelt (e.g., "abscence", "accomodation", "fourty"), including proper nouns (e.g, "Chronbach", "Bronfennbrenner"), or the wrong word was used (e.g., "existenting", "illuded", "meaningly"). I accepted either British or American English spellings and ignored scientific words and jargon terms. Notably, the process of converting PDFs to text introduced some errors in the text, such as "aefct" or "efcfiacy" instead of "affect" and "efficacy". While I corrected these instances in the titles and abstracts, the broader vocabulary of full-texts was too large to similarly clean. As such, I examined only the abstracts and titles for spelling errors. I used $X^2$ tests to compare each group's proportion of titles and abstracts with spelling errors and, where I observed differences, pairwise comparisons of proportions to identify which groups differed.

*3.3.3. Readability*

I calculated readability scores for the abstracts and full-texts of the articles using the Flesch Reading Ease score (FRE) in the sylcount package (Schmidt, 2022). The FRE compares the number of syllables to the number of words in the text, and the number of words to the number of sentences (Kincaid et al., 1975), as shorter words and sentences are more readable. Most texts score between 10 and 100, with 60-70 indicating a United States 8th grade reading level. However, academic articles are typically more complex and aimed at college-level readers and thus associated with scores lower than 50.

*3.3.4. References and citations*

I calculated four indicators related to references and citations: i) the number of references for each article, as extracted by JATSdecoder, ii) the number of citations received by each article, iii) the percentage of references that were self-references, and iv) the percentage of citations that were self-citations. Due to the non-normal distributions of these variables, for these indicators I used the Kruskal-Wallis H test for differences between groups and, when differences were observed, the Wilcox pairwise tests to identify which groups varied from one another.

I obtained from OpenAlex[4] as of August 2023 the number of citations each article had received since publication and the 3 year citations for articles published before 2020. I matched sampled articles to OpenAlex based on DOI. I used OpenAlex as it has extensive coverage and was more recent than the KB's Dimensions database. The match rate was 84.8% of QJ articles, 68.6% of WoS articles, and 90.2% of mid-tier articles. The lower coverage of WoS articles stemmed largely from two journals whose articles could not be matched.

The other two indicators related to self-references – the percentage of an article's references that were to any of the authors' previous articles – and self-citations – the percentage of an article's citations that were from any of the authors' later articles. To match authors of the sampled articles with citing and referenced authors and articles, I retrieved the author and article IDs of the sampled, citing, and referenced articles/authors from Semantic Scholar – a free index of more than 200 million academic articles[5] – using the semscholar package (Jahn, 2023). The coverage of sampled articles in Semantic Scholar was 96.7% of QJ articles, 92.6% of WoS articles, and 91.9% of mid-tier articles. I identified the articles referenced by a sampled article based on the article ID, and then identified self-references by matching the author IDs between sampled and referenced articles. The percentage of self-references was derived from the proportion of each article's referenced articles that were published by any of the sampled article's authors. Similarly, I identified self-citations by matching citing articles and authors with the sampled article's ID and author IDs. The percentage of self-citations was derived from the proportion of each article's total citations that stemmed from an article published by any of the sampled authors.

*3.3.5. Size and internationality of authorship teams*

I also used the author IDs gathered from Semantic Scholar to identify the number of authors of each article. I extracted the list of authors' countries for each article using JATSdecoder then counted the unique number of countries per article. I used a Kruskal-Wallis H test to compare the groups due to the non-normal distribution of the number of authors and countries involved, followed by Wilcox pairwise tests to identify which groups differed.

*3.3.6. Ethics approval and informed consent*

As the articles sampled were all published in psychology journals, it could be assumed that the majority of studies using human participants should acknowledge that they received or were exempt from seeking ethics approval and/or received informed consent from their participants. To assess the prevalence of mentions of ethics approval in studies that included human participants, I searched the full-texts of articles for the word "participant" used in combination with at least one of the following terms: review board, ethic/ethics/ethical committee/commiete, committee(s) on (the) ethics, ethics commission, ethics screening committee, ethic(s) board, ethical clearance, ethics statement, ethical/ethics standards, ethical/ethics guidelines, or ethical/ethics approval. Similarly, to identify articles that mentioned informed consent, I searched the full-texts of articles for the word "participant" used in combination with informed consent, consented, or consent. The terms for ethics and consent were identified by reviewing a subset of the sampled articles and collating a list of relevant terms. I used $X^2$ tests to compare each group's proportion articles that mentioned these concepts and, where I observed differences, pairwise comparisons of proportions to identify which groups differed.

This process should capture the majority of articles that mention participants and ethics or consent in a standard way. However, this process is not infallible. For instance, the sampled articles may use "participants" in regard to a prior study's methods or results, state that they did not use participants in the study, or may have used a different term for ethics/consent. As such, these results may underestimate the prevalence of mentions of ethics and informed consent in the sampled articles.

*3.3.7. Incorrectly reported p-values*

I used the statcheck package (Nuijten & Epskamp, 2023) to identify incorrectly reported p-values. Statcheck extracts test statistics reported in American Psychological Association (APA) format, recalculates the p-values based on the values reported by the authors and compares the recalculated and reported p-values to identify instances of misreporting. Statcheck accounts for rounding and distinguishes between cases in which the misreporting results in an erroneous decision about the statistical significance of a finding. For instance, if a study reports a p-value indicating a finding is statistically significant but statcheck's recalculated p-value indicates a non-significant finding (or vice versa), such a case would constitute an erroneous decision. In this study, I include only these erroneous decision cases. Further, although multiple instances of misreporting may occur in an article, I measure here only the overall number of articles in each group that contained at least one misreported p-value and as a percentage of those that reported at least one APA test statistic. I used $X^2$ tests to compare each group's proportion articles with misreported p-values and, where there were differences, pairwise comparisons of proportions to identify which groups differed.

---

[4] https://docs.openalex.org/download-all-data/openalex-snapshot
[5] https://www.semanticscholar.org/about

APA is a common reporting style in the psychological sciences and thus is appropriate for detecting statistical information in the sampled articles. However, articles that use formatting inconsistent with the APA style cannot be assessed, nor can articles in which the author has incorrectly or incompletely reported the information required to recalculate the p-value. The conversion of PDFs to text files may also have affected how results were presented and their ability to be assessed by statcheck. As such, these results likely underestimate the prevalence of errors in p-value reporting in the sampled articles.

## 4. Results

The study analysed 3 samples: 1,714 articles in 31 QJs from 22 publishers; 1,691 articles in 16 WoS-indexed journals from 15 publishers; and 1,900 articles in 45 mid-tier journals from 31 publishers. Regarding the length indicators, there were significant differences between groups in both the lengths of abstracts ($F(2, 5295) = 23.7, p<0.00$) and of full-texts ($F(2, 5302) = 134.8, p<0.00$; panels A and B in Figure 2). Abstracts in QJs averaged 186 words and were significantly shorter than those in WoS journals (196 words, p<0.00) or mid-tier journals (201 words, p<0.00). Abstracts in WoS and mid-tier journals were similar in length (p=0.06). Full-texts of QJs were on average 4,229 words and also significantly shorter than full-texts in WoS (5,434, p<0.00) or mid-tier journals (5,143, p<0.00). WoS journal articles were also significantly longer than mid-tier journal articles (p<0.00).

Regarding readability of the full-texts, the one-way ANOVA indicated there was a significant difference between groups ($F(2, 5289) = 10.3, p<0.00$), with mid-tier articles slightly less readable than the other groups (p<0.00). However, the median scores for all groups ranged between 42.7 and 43.9 – the expected range for academic texts – suggesting there is limited practical difference between groups (panel C, Figure 2). Further, these median scores are all below 50, which is appropriate for an academic article aimed at university-educated readers.

The percentage of abstracts that contained spelling errors was low in all groups, at 2.2% or less (panel D, Figure 2). However, there were significant differences between groups ($X^2 (2, N = 5,305) = 10.4, p<0.01$), with a greater likelihood of errors in QJ's abstracts than other journal types (p<0.05). The proportion of titles that contained spelling errors was comparable between groups ($X^2 (2, N = 5,305) = 4.4, p=0.11$).

Figure 3 shows for each group the results of indicators related to references and citations. The number of items referenced by the sampled articles differed significantly between groups (F(2, 5259) = 166.7, p<0.00; panel A). Articles in WoS journals contained significantly more references (mean = 50.3) than the other two groups (p<0.00), and mid-tier articles had more references (mean = 46.3) than articles in QJs (mean = 34.6, p<0.00).

The journal groups also significantly differed in the number of citations received within 3 years of publication (H (2) = 482.1, p<0.00; panel B) and between publication and August 2023 (H (2) = 744.5, p<0.00, panel C). In both cases, articles in WoS journals received significantly more citations than the other journal types (3 year mean = 3.8, all time mean = 16.7, p<0.00), and articles in QJs (0.8, 3.3) received significantly fewer citations than mid-tier journals (2.4, 12.4, p<0.00).

There were significant differences between groups in the percentage of references that were references to articles previously published by any of the authors of the sampled articles (H (2) = 216.8, p<0.00; panel D), and also in the percentage of citations that stemmed from any of the authors' later articles (H (2) = 76.5, p<0.00; panel E). Authors of articles in WoS journals had the most self-references (mean = 6.8%, p<0.00), and mid-tier journals' references contained more self-references (mean = 5.6%) than QJ articles (mean = 4.2%, p<0.00). The same pattern arose in self-citations, with WoS journals having the highest levels of self-citations at on average 17.6% (p<0.00), and mid-tier journals (mean = 16.2%) having higher levels than QJ articles (mean = 15.9%, p<0.01).

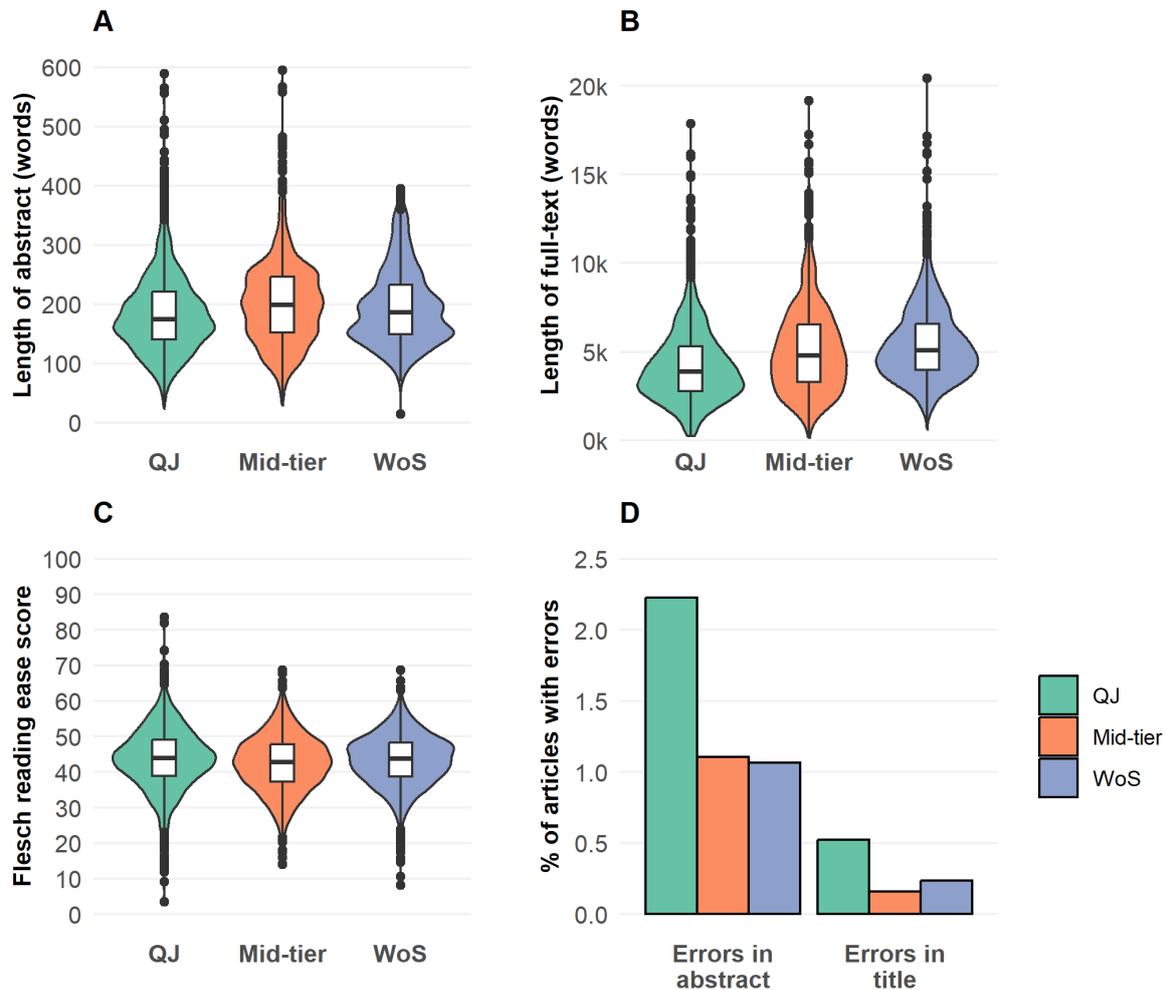

**Fig. 2** Distributions of the lengths of abstracts (A) and full-texts (B), the FRE scores (C), and the percentage of abstracts and titles with spelling errors (D) in each group.

The distributions of the number of authors and countries involved in the articles of each journal group are shown in Figure 4. There were significant differences between groups in the number of authors (H (2) = 240.3, p<0.00), with articles in QJs involving significantly fewer authors (mean = 2.7) than the other journal types (p<0.00). WoS journals also had significantly more authors (mean = 3.8) than mid-tier journals (mean = 3.3, p<0.00). There were also significant differences between groups in the number of countries involved in articles (H (2) = 107.6, p<0.00): articles in QJs were less often international collaborations (mean = 1.1) than mid-tier and WoS journals, which each averaged 1.3 countries (p<0.00), although the similar means and interquartile ranges indicate limited practical difference.

Before calculating the percentage of articles the mentioned ethics and informed consent, I first identified the articles that included participants. Just over two-thirds of articles in QJs had participants (1,195, 69.7%), compared to 78.4% of articles in mid-tier journals (1,489) and 89.8% of WoS journals (1,519). The percentage of these articles that mentioned ethics and informed consent are shown in panels A and B of Figure 5.

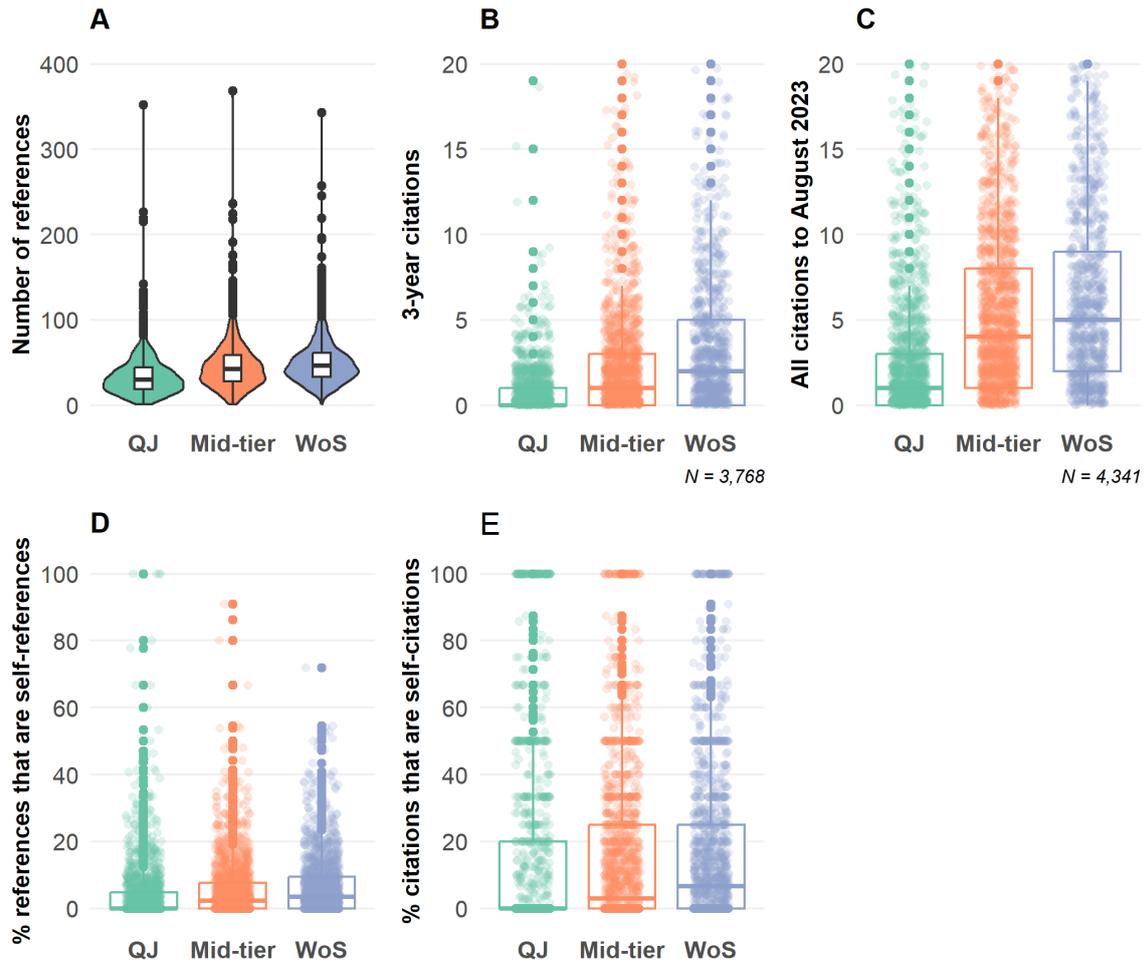

**Fig. 3** Distributions of the number of references (A), the number of 3-year citations (B), the number of citations received between publication and August 2023 (C), the percentage of references that are self-references (D), and the percentage of citations that are self-citations (E).

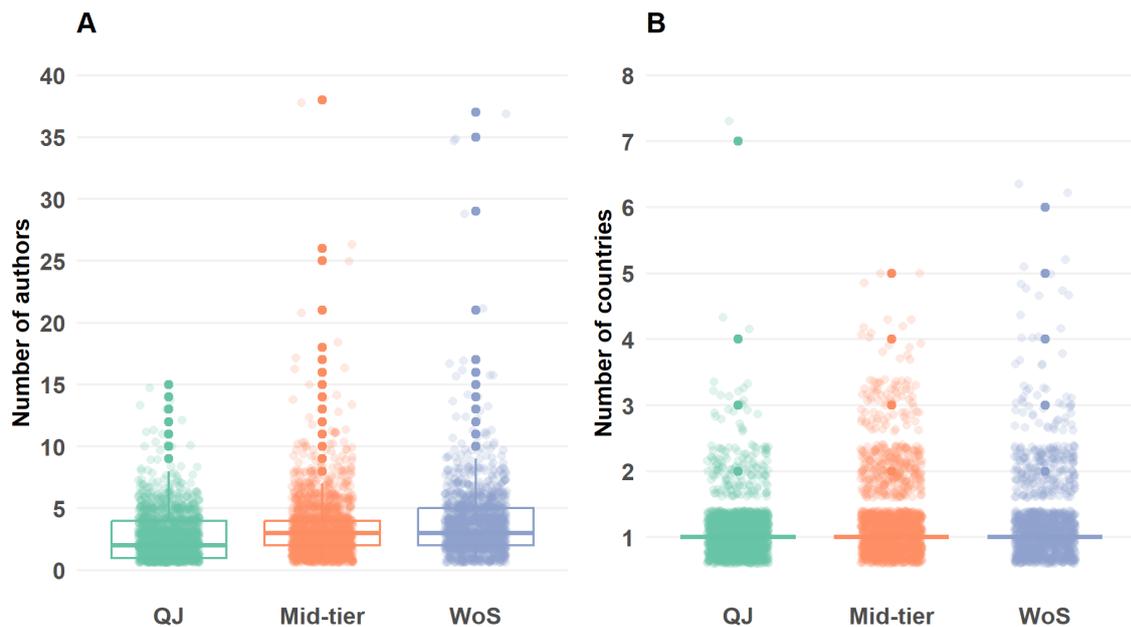

**Fig. 4** Distributions of the number of authors (A) and number of countries (B) involved in articles published in QJ, mid-tier journals and WoS-indexed journals.

A 3-sample test for equality of proportions identified significant differences between groups both in the proportions of articles that mentioned ethics ($X^2$ (2, N = 1,533) = 133.3, p<0.00) and consent ($X^2$ (2, N = 2,187) = 83.6, p<0.00). Within each group, articles were more likely to address informed consent than they were to mention ethics. Only 25.7% of QJ articles mentioned ethics, which was significantly less than the other two groups (p<0.00). Mid-tier articles were also less likely to mention ethics (34.5%) than WoS journals (46.9%, p<0.00). Similarly, WoS journal articles were more likely to mention consent (61.2%) than either QJ articles (44.9%) or articles in mid-tier journals (48.4%, p<0.00). The difference between mid-tier and questionable journals was not significant (p=0.09).

Panel C of Figure 5 shows the percentage of articles that reported the results of at least 1 statistical test in APA format. This included 395 articles in QJs (23.0%), 442 articles in mid-tier journals (23.3%), and 551 articles in WoS journals (32.6%). Panel D shows the percentage of these articles that made decision errors about the significance of their findings due to misreported p-values: 7.9% of mid-tier journals, 10.5% of WoS journals, and 12.7% of QJs. These percentages were not statistically different ($X^2$ (2, N = 1,388) = 5.1, p=0.08). However, one must consider in interpreting these results that less than a third of articles in each group reported a statistical test and were thus able to be considered in this analysis.

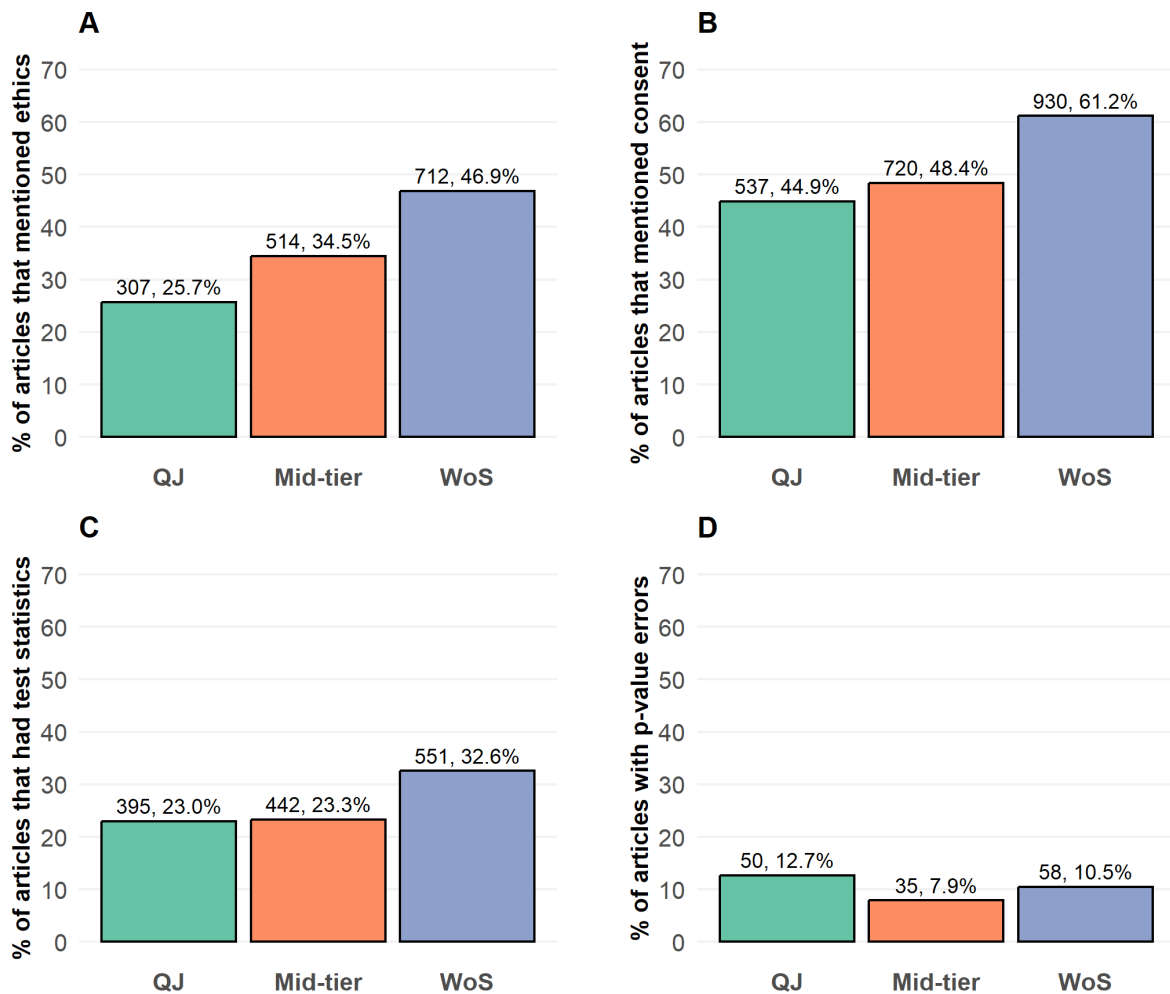

**Fig. 5** The percentage of articles with participants that mentioned ethics (A) and consent (B), and the percentage of articles with assessable test statistics (C) that made errors in significance decisions due to misreported p-values (D).

## 5. Discussion

This study compared samples of articles in QJ, mid-tier, and WoS-indexed journals on several quantitatively assessable indicators of quality relating to theoretical embedding, methodology, communication, impact, and collaboration. Broadly, these indicators can be mapped to Langfeldt et al's (2020) concepts of space-type value (citation impact) and the plausibility or reliability of the study in its theoretical grounding, methods, and communication. The aims of the study were to determine whether articles in QJs were distinguishable from those in non-QJs based on these indicators of quality, and identify what can consequently be deduced about the quality of articles in QJs.

Differences emerged between all groups for most indicators and in a pattern that could be expected based on the spectrum of quality originally assumed: WoS-indexed journals scored in the direction most indicative of high quality and QJs scored lowest. As such, QJ articles and abstracts were significantly shorter, referenced significantly less literature, received significantly fewer citations, involved significantly fewer authors, and mentioned ethics and consent processes significantly less often than articles in WoS and mid-tier journals. These differences might reflect an absence of editorial or peer review processes – a common feature of QJs. We have previously observed the length of manuscripts and the number of references to increase following peer review as reviewers request additional methodological details, interpretation of results, and related literature (Stephen, 2022; Akbaritabar, Stephen & Squazzoni, 2022). Peer reviewers and editors may also identify spelling errors for correction and request ethics and consent processes be detailed. Further, higher quality journals may have policies regarding the inclusion of ethics and consent statements for relevant studies, as is the case for Sage journals[6], for example, which increases the likelihood these details are reported. However, only 50-60% of WoS indexed articles with participants mentioned ethics and consent, suggesting under-reporting of these processes is not a problem limited to mid-tier and questionable journals. A lack of peer review and editorial oversight may thus explain the poorer performance of QJs on these indicators of length, referencing, and ethics and consent statements.

QJs also received significantly fewer citations than mid-tier and WoS journals. While this may suggest these articles are of less utility to other researchers, it may also reflect that these journals are less visible than mid-tier and WoS journals as they are not included in WoS, DOAJ, or other key indices. Notably in relation to citations, articles in WoS journals contained a higher percentage of references that were to the authors' own previous work and a higher percentage of citations were self-citations. While self-citing is a common and generally acceptable practice in academia to build upon previous work, the higher levels of this practice by authors in WoS articles – in particular self-citations which accounted for nearly 20% of all citations – exceed the disciplinary standard set by mid-tier and QJ articles. Also notable is that in each group a similar percentage of articles that reported statistical tests made erroneous decisions due to misreported p-values. For other indicators, there were statistically significant differences between groups, but the practical implications were limited, e.g., the readability of articles and internationality of authorship teams, which, while lower in the mid-tier and QJ articles respectively, was similar in all groups.

## 5.1. Limitations

The following limitations should be considered in relation to this study. First, reading PDFs inherently introduces some error into the data due to mistakes in optical character recognition or unexpected formatting. While cleaning was undertaken for the abstracts and titles, this could not be conducted for the full-texts due to the size of the corpus. As such, the prevalence of mentions of ethics and informed consent and erroneous decisions due to misreported p-values may be underestimated, and estimates of readability and full-text length may be influenced should words have been inadvertently merged or separately. However, as the same process as applied to all articles, no particular set was (dis)advantaged by any errors introduced and so comparing between groups remains suitable.

Secondly, journal policies may have influenced the lengths of abstracts and full-texts, using APA formatting and thus reporting statistics in a detectable format, and the requirement to report ethics and consent statements. Such effects are potentially evident in the sharp cut-off of abstracts at approximately 400 words in WoS-indexed journals, for instance. Due to the relatively large number of journals involved (92), I have not accounted for variations in such policies and thus cannot estimate their effect on the results.

---

[6] https://uk.sagepub.com/en-gb/eur/publication-ethics-and-research-integrity-policy-guidelines-for-authors

Finally, in assessing quality, quantitative indicators cannot replace a qualitative review of articles to evaluate some aspects of quality, such as the suitability of methods for the study's aim, the validity of conclusions drawn based on the study's results, or the original contribution of the study to the field. As such, I was unable here to assess articles in terms of their originality, under Langfeldt et al's (2020) framework, or other important aspects of quality. However, the use of automated, quantitative processes here has allowed for an objective analysis of several indicators associated with quality in a large sample of articles.

*5.2. Conclusions*

In summary, these findings suggest that QJ articles do diverge from the disciplinary standards set by peer-reviewed journals in psychology on quantitative indicators of quality that tend to reflect the effect of peer review and editorial processes. In line with Langfeldt et al's (2020) framework, these indicators suggest a lower plausibility of articles in QJs than other journals. In addition, articles in QJs are valued less – as represented by citations – than WoS and mid-tier journals. However, this may occur due to the lower visibility of QJs. Notably, mid-tier and WoS journals are also affected by potential quality concerns, such as under-reporting of ethics and informed consent processes, and the presence of errors in interpreting statistics.

This study presents an initial examination of the feasibility of measuring article quality via quantitative indicators and drawing subsequent conclusions about the quality of articles in QJs. Much research remains to be done in this space to ensure we have a comprehensive understanding of the potential issue QJs represent in contaminating academic discourse with poor quality content, so that we may respond appropriately and proportionately. This includes further qualitative research into the quality of articles in QJs, and also how we classify journals as questionable, accounting for geopolitical influences and the effect our research evaluation systems have on researcher publishing practices.

**Statements and declarations:**

Competing interests: I have no competing interests to declare.

Funding: This research did not receive any specific grant from funding agencies in the public, commercial, or not-for-profit sectors.

Data statement: As the journals included were defined by data from the proprietary WoS and Cabell's databases, I cannot make these datasets publically available.